\documentstyle[12pt,epsfig]{article}
\begin{document}

\hfill FTUV/96-11, IFIC/96-12

\hfill HU-TFT-96-9

\hfill hep-ph/9602387

\vspace*{1cm}

\begin{center}

{\bf \Large  Single W-boson production in $e^-\gamma$ colliders}
\footnote{Talk given by M. Raidal in the  workshop "Physics with 
Linear Colliders," Gran Sasso, 1995.}

\vspace*{0.3cm}

{\large K. Huitu$^a$, J. Maalampi$^b$ and M. Raidal$^c$}  \\

\vspace*{0.3cm}
 
{\it \small $^a$Research Institute for High Energy Physics,
University of Helsinki \\
$^b$Department of Theoretical Physics,University of Helsinki \\
$^c$Department of Theoretical
Physics, University of Valencia
}

\end{center}

{ \small Single $W$-boson  production in $e^-\gamma$ collisions with
polarized beams is investigated.
In the framework of the Standard Model the
updated estimates for the measurement precision of
 photon anomalous coupling parameters $ \kappa_{ \gamma},$ 
$ \lambda_{ \gamma}$ at the Next 
Linear Collider with $ \sqrt{s_{ e\gamma }}=420$ GeV are obtained.
The production of right-handed gauge bosons $ W^-_{ 2 }$ in this
collision mode is also analysed. If the associated neutrino is light,
the channel would give the best discovery reach for $ W^-_{ 2 }$
in the Next Linear Collider.
}
\normalsize

\baselineskip=18pt

\vspace*{1cm}
%TEXT BEGINS HERE

% DEFINE YOUR MACROS, etc.. HERE

\newcommand{\be}{\begin{equation}}
\newcommand{\ee}{\end{equation}}

\newcommand{\bea}{\begin{eqnarray}}
\newcommand{\eea}{\end{eqnarray}}

\newcommand{\bean}{\begin{eqnarray*}}
\newcommand{\eean}{\end{eqnarray*}}

\newcommand{\ka}{\kappa_{ \gamma}}
\newcommand{\la}{\lambda_{ \gamma }}

\newcommand{\ba}{\begin{array}}
\newcommand{\ea}{\end{array}}

\newcommand{\bchi}{\bar{\chi}}
\newcommand{\bG}{\bar{G}}
\newcommand{\BG}{{\bf g}}
\newcommand{\BH}{{\bf h}}
\newcommand{\cD}{{\cal D}}
\newcommand{\cL}{{\cal L}}
\newcommand{\al}{\alpha}
\newcommand{\B}{\beta}
\newcommand{\de}{\delta}
\newcommand{\DE}{\Delta}
\newcommand{\g}{\gamma}
\newcommand{\ep}{\epsilon}
\newcommand{\vep}{\varepsilon}
\newcommand{\th}{\theta}
\newcommand{\LM}{\Lambda}
\newcommand{\lm}{\lambda}
\newcommand{\vp}{\varphi}
\newcommand{\ie}{{\it i.e. }}
\newcommand{\eg}{{\it e.g. }}
\newcommand{\LB}{| \! [}
\newcommand{\RB}{] \! |} 
\newcommand{\lr}{{left-right symmetric model}}

\section{Introduction}

In addition to the  electron-positron option, 
the electron-electron and electron-photon collision modes 
of the Next Linear Collider (NLC) are also
technically realizable  \cite{orava}. 
During the recent years the physics potential of
the  latter options has
been under intense study.  While $e^-e^-$ collisions
have been found to be particularly suitable for
the study of possible  lepton number violating 
 phenomena \cite{cuy},  the $e^-\gamma$ operation 
mode will also be well motivated
from the  point of view of new physics.

So far, the $ e^-\gamma$ collisions have been studied using 
the photon spectrum
of  classical Bremsstrahlung. 
In the linear collider it will be possible to obtain high luminosity 
photon beams by  backscattering intensive 
laser pulses off the electron beam
\cite{ginz} without considerable losses in the  beam energy and with
very high polarizability and monochromaticity \cite{telnov}.     
 This possibility makes the  $ e^-\gamma$ collisions  
 ideal for studying heavy gauge boson
production processes \cite{eg,choi,king}, 
since the initial state photon provides us with a possibility
 to probe directly  the gauge boson self-interactions.

 We will consider a single massive vector boson production in
$ e^-\gamma $ collision,  
\be
e^-\gamma\rightarrow W^-N,
\label{prot}
\ee
for any combination of beam polarization.
Here $ W^-$ may stand for the ordinary SM charged vector boson
$ W^-_1$ and $ N$ for the massless Dirac 
electron neutrino $ \nu_e$. However, we do not restrict ourselves only
to this case, since a wide class of models beyond 
the SM predicts a existence of
new heavy vector bosons and massive neutrinos. For example,
in the left-right symmetric model (LRM) \cite{lr} 
the vector boson may  also be a heavy  right-handed 
weak boson $ W_2$.  
The present lower limit for the mass of $ W_2$ coming from high energy
experiment is
$M_{W_2}\geq 652$ GeV \cite{mass}, so that the 
right-handed boson production
will be kinematically forbidden at least 
in the initial phase of the  NLC. 
At the final phase
of the NLC, however, the reaction (\ref{prot}) 
may be kinematically allowed and  even
favoured compared with, \eg, the  $ W_2$ pair production in $ e^-e^+$
collisions, since the mass of
 the  associated neutrino   could be smaller 
than the mass of  $ W_2.$ In the case of a sizeable     
mixing  between the light, predominantly left-handed and the heavy,
predominantly right-handed  neutrinos the study of
the  process (1)  may extend the kinematical 
discovery range of $ W_2 $ almost up to the energy 
$ \sqrt{s_{ e\gamma} }.$

\section{Anomalous triple boson coupling in the Standard Model}

 There are two Feynman diagrams 
 contributing at the tree level to the reaction (\ref{prot}) (see Fig. 1).
One of them, the t-channel diagram, 
involves a triple gauge boson coupling  
making the  process  suitable for testing the non-Abelian gauge
structure of the theory. A particularly interesting feature of the process 
(1) is that it is sensitive  only to the possible  
anomalous coupling of the  photon,
allowing one  to discriminate between the photon anomalous coupling 
and the  anomalous coupling of  massive
neutral gauge boson $ Z^0.$ 
Since the  deviation  from the SM coupling is expected to be small, 
one can use of polarization  of the initial state particles to
enhance  these effects.  

The most general $ CP$-conserving 
$\gamma WW$ interaction allowed by the electromagnetic gauge invariance
is of the form \cite{hagi}
\bea
{\cal L}_{ \gamma WW } & = & -ie(W^{\dagger}_{ \mu\nu }W^{\mu}A^{\nu} -
W^{\dagger}_{ \mu }A_{\nu}W^{\mu\nu} + \kappa_{ \gamma }
W^{\dagger}_{ \mu }W_{\nu}F^{\mu\nu} + \nonumber \\
& & \frac{\lambda_{\gamma}}{M^2_W}
W^{\dagger}_{ \tau\mu }W^{\mu}_{\nu}F^{\nu\tau}),
\label{triple}
\eea
where $ W_{ \mu\nu }=(\partial_{ \mu } - ieA_{ \mu })W_{ \nu } - 
(\partial_{ \nu } - ieA_{ \nu })W_{ \mu }$ and
$ F_{ \mu\nu }=\partial_{ \mu }A_{ \nu } - \partial_{ \nu }A_{ \mu }.$
The coefficients $ \kappa_{ \gamma }$ and $ \lambda_{ \gamma }$ are 
related to the magnetic moment $ \mu_W$ and the electric quadrupole 
moment $ {\cal Q}_W$ of  $ W$ according to
\[\mu_W=\frac{e}{2M_W}(1 + \kappa_{ \gamma } + \lambda_{ \gamma }), 
\;\;\;\;\;\;\;\;\;\; {\cal Q}_W=-\frac{e}{M^2_W}(\kappa_{ \gamma } -  
\lambda_{ \gamma }).\]
In a gauge theory at  tree level the coefficients have the values
$ \ka=1$ and $ \la=0$.

In the context of 
the SM, the  process (1)
has already been investigated previously 
 \cite{eg,choi,yehudai,phil}, and its 
sensitivity to the photon 
anomalous coupling has been found to be comparable      
with the estimated sensitivity of 
the $ W$ pair production processes \cite{choi}.  
 We have  updated these  analysis 
 for a  $ 500$ GeV $ e^+e^-$ collider, 
taking into account the effects of beam  polarization
 and final state polarization as well as
 the recent developments in the linear collider design.
The expressions for the  general 
helicity amplitudes of the process are given in
ref.\cite{raidal}.
For determination the cross sections from the helicity amplitudes
we have assumed 100\% longitudinally 
 polarized  electron and linearly polarized photon beams. This is
 an often used approximation since  
in practice the polarizations will  be more than 90\%.

The scattering of linearly polarized laser light off the electron beam
produces a polarized photon beam with very 
hard spectrum strongly peaked at 
the maximum energy, which is about $ 84$\% of the electron beam energy
\cite{ginz}. There are two different collision schemes of the photon colliders
possible \cite{telnov}. In the first case the photon conversion region is very
close to the interaction point and  the 
entire photon spectrum interacts with the 
electron beam. From the physics point of view this realization of the
$ e^-\g$ collisions is undesired because of the high rate 
of the background processes initiated 
by the electrons which have been  used for creating the photon beam and 
also because of the low monochromaticity of the photon beam.

In the case of the second collision scheme the
distance  between the conversion and interaction points is longer. 
The electrons used for producing the photon beam are removed 
by applying  strong magnetic field
 and therefore the  $ e^-\g$ collisions are  clean.
Since the
electron beam probes only the     
hardest photons of the $ \g$ beam the
collisions are highly monochromatic.
 The achievable luminosities
in this case are found to vary from 30 fb$^{-1}$ at VLEPP to 200 fb$^{-1}$
at TESLA per year \cite{telnov}. 
Therefore, we have  carried out our analysis
for the center of mass energy $ \sqrt{s_{ e\g }}=420$ GeV 
corresponding to the peak value of the photon spectrum,  
assuming that the relatively 
small nonmonochromaticity effects of the photon beam as well as the
nonmonochromaticity effects of the electron beam due to the 
energy losses in beamstrahlung ( both at the level of a few percent)
 will be taken into account in the analysis of  experimental data.
The other relevant  NLC parameters which we 
have used  are the following: 
 integrated luminosity  $ L_{ int }=50$ fb$^{-1},$
 the covering region of a detector  $ |\cos\th|\leq 0.95$
and  $ W^-$ reconstruction
efficiency of $ 0.1.$

We have used five observables 
for testing the parameters $ \ka$ and $ \la.$ 
Obviously, the differential cross section $ d\sigma_{ \tau_1=\pm1 }/
d\cos\theta$
and the total cross section 
$ \sigma_{ \tau_1=\pm1 }^{tot}$ for different 
photon beam polarization $ \tau_1=\pm1$ 
can be analysed.
Since  the differential cross sections are 
strongly peaked in the backward
direction one would expect that also the forward backward asymmetries
\be
A^{ FB }_{\pm}=\frac{\sigma_{ \tau_1=\pm1 }(\cos\theta\geq 0)- 
\sigma_{ \tau_1=\pm1 }(\cos\theta\leq 0)}
{\sigma_{ \tau_1=\pm1 }(\cos\theta\geq 0) +
\sigma_{ \tau_1=\pm1 }(\cos\theta\leq 0)} 
\label{fb}
\ee
could be sensitive to the anomalous coupling.
The quantity, which reflects the effects of
the beam polarization, is the polarization asymmetry
$ A_{ pol }$ defined as
\be
A_{ pol }(\cos\theta)=\frac{d\sigma_{ \tau_1=+1 }- 
d\sigma_{ \tau_1=-1 }}
{d\sigma_{ \tau_1=+1 } +
d\sigma_{ \tau_1=-1 }}.
\label{as}
\ee
We have also studied whether the measurement of the final state
$ W$-boson polarization could offer 
sensitive tests for $ \ka$ and $ \la.$
The information about the polarization of $ W$-boson can be obtained
by measuring the angular distribution of its decay products.
A suitable quantity would be 
the forward-backward asymmetry of the leptons produced in  $ W^-$
decay, which  is related to the cross 
sections corresponding to the different 
$ W^-$ polarization states $ \tau_2=\pm1$
as follows (see \eg ref.\cite{yehudai}):
\be
\chi_{ \pm }^{FB}=\frac{3}{4}\frac{\sigma_{\tau_1=\pm1}^{\tau_2=-1} -
\sigma_{\tau_1=\pm1}^{\tau_2=+1}}{\sigma_{\tau_1=\pm1}^{tot}}.
\label{xfb}
\ee
We have carried out  a $ \chi^2$ analysis   
by comparing the SM prediction of the observables  with
those corresponding ones to the anomalous $ \ka$ and $\la.$ 
The limits
are calculated at $ 90$\% confidence level, which corresponds to
$ \Delta\chi^2=4.61.$
The statistical errors are computed assuming the NLC parameters
given above. The systematic errors are estimated by assuming 
the  uncertainty of the cross section measurement to be
at the level of $ \sim2$\% \cite{sys}, coming mainly from the
errors in the luminosity measurement, the acceptance, 
the background subtraction 
and the knowledge of branching ratios.

Both forward-backward asymmetries, $ A_{ \tau_1 }^{FB}$ and 
$ \chi_{ \tau_1 }^{FB}$, turned out to be several times
less sensitive to the anomalous 
coupling than the  other three observables.
Polarization asymmetry $ A_{ pol }$  
and the total cross sections 
$ \sigma_{ \tau_1=\pm1 }^{tot}$ are more sensitive 
to the deviations from the SM but still do not  allow 
to constrain couplings sufficiently. 
The most sensitive observable to the photon anomalous coupling
is the differential cross section.
The contours of  allowed regions in $ (\ka,\la)$ space
 obtained from its analysis  are plotted in Fig. 2.
The curves for the different photon polarization states   
$ \tau_1=\pm1$ are indicated in the figure. 
The contour resulting from the combined analysis is denoted by $ a.$
As can be seen from Fig. 2 the most stringent 
constraints for the anomalous
coupling are obtained in the case of  left-handedly polarized
electron and right-handedly polarized photon beams. 
This is an expected result, since the s-channel diagram in Fig. 1 
does not contribute in this case and the entire  cross section
comes from the t-channel diagram, 
which probes the triple boson coupling.

As a result,  by studying the reaction (1) in the NLC
with  the assumed set of parameters,
one could  constrain the anomalous triple boson coupling
 parameters $ \ka$ and $ \la$ to the following regions:
\[ -0.01\leq 1-\ka\leq 0.01, \;\;\;\;\;\;\;\;\
 -0.012\leq\la\leq 0.007 .\]
At this level of  precision the radiative corrections are expected
to start to play a role \cite{rad}.
Since the size of the SM radiative corrections depends crucially on the 
beam polarization, the use of polarized beams allows one 
to discriminate between 
the the radiative corrections and corrections from the new physics.

We have repeated the analysis with the integrated luminosity of 10$fb^{-1}$
which gives $\sim 1.4$ times weaker bounds for parameter $ \ka$ and 
$\sim 1.6$ times
weaker bounds for parameter $ \la.$ This shows that the measurement 
uncertainties are largely dominated by the systematic errors.  
In order to see the relevance of the beam polarization 
and to compare our results with the earlier works with unpolarized beams 
we also repeated  the analysis using  the NLC parameters of 
ref.\cite{choi} \ie  $ |\cos\th|\leq 0.7$ and the integrated luminosity
of 10$ fb^{ -1 }$. It turned out that the beam polarization (together
with the monochromaticity of the photon beam) gives an improvement of
a factor of $3$ in the measurement precision of the anomalous coupling 
parameters $ \ka$ and $ \la.$

\section{Single heavy vector boson production in  left-right model}

The LRM \cite{lr} is an extension of the
SM, in which  the gauge interactions of left-handed
and right-handed fundamental fermions are treated on equal basis.
  The LRM is based on the gauge symmetry 
$SU(2)_R\times SU(2)_L\times U(1)_{B-L}$, and
there are hence two new weak bosons, $W_2$ and $Z_2$,  in addition to
the ones known in the SM. The  left-right symmetry, not present  in
the low energy world, is broken by a SU(2)$_R$ triplet Higgs field
$\Delta=(\Delta^{++},\Delta^+,\Delta^0)$. The only new fermions the
model predicts are the right-handed neutrinos.
 
The energy scale
$v_R=<\Delta^0>$ of the breaking of the LRM  symmetry to the SM
symmetry, which also sets, up to coupling constants, the mass scale
of the  new weak bosons and right-handed neutrinos, is not given by
the theory itself. 
In the Tevatron one has made a direct search of
$W_2$ in the channel $pp\to W_2\to eN$. The bound they give is
$M_{W_2}\ge 652$ GeV \cite{mass}. The result is based on several
assumptions on the LRM: the quark-$W_2$ coupling has the SM strength,
the CKM matrices for the left-handed quarks and the right-handed
quarks are similar and the right-handed neutrino does not decay in the
detector but appears as missing $E_T$. If one relaxes the  first two
assumptions, the mass bound will be  weakened considerably, as was
pointed out in ref.\cite{rizzoap95}. The third assumption is also
crucial; if the right-handed neutrino is heavy, with a mass of say
100 GeV or more, it will decay in the detector  into charged
particles with no missing energy. For this case, which is natural
in  the LRM,  Tevatron search would be ineffective. 
Therefore, it was argued in ref.\cite{rizzoap95}, 
that the lower limit for $M_{W_2}$ could be as low
as 300 GeV.

  The mass dependence of the  total cross section of the process $
e^-_R\g\rightarrow W^-_2 N$ can be seen in Fig. 3, where we plot the
cross section as a function of
$ W^-_2$ mass for the center of mass energy
$ \sqrt{s_{ e\g} }=1.5$ TeV, expected to be possible to achieve in
the final stage of NLC, assuming the left- (Fig. 3 (I)) and
right-handedly (Fig. 3 (II)) polarized photon beams. The curves
denoted by $ a$ and $ b$ correspond to the neutrino masses
 $ M_N=300$ GeV and $ M_N=600$ GeV, respectively. The cross sections
are found to be reasonably large for almost the entire kinematically
allowed mass region, decreasing faster with $ M_{ W_2 }$ for the $
\tau_1=1$ photons. At low $ W_2$ masses the difference between $ a$
and $ b$ curve is small but for heavy $ W_2$ masses the cross section
depends strongly  on the neutrino mass. If $ M_N\leq M_{ W_2 },$ 
the reaction (1) enables us to study heavier vector bosons than what is
possible  in the $ W^-_2$ pair production in
$ e^-e^+$ or $ e^-e^-$ collisions. 
 
The reaction  would be even more useful in this respect
  if the mixing between the heavy and the light neutrino  is large
enough to give observable effects. In Fig. 4 we plot the cross
section of the  reaction
$ e^-_R\g\rightarrow W^-_2 \nu$ for different photon polarizations
assuming a vanishing  mass of $ \nu$ and the neutrino mixing angle of
$ \sin\theta_N=0.05.$   For this set of parameters the process should
be observable  up to $ W$-boson mass $ M_W=1.2$ TeV.

\section{Summary} 

We have studied  usefulness of the reaction $e^-\gamma\to W_2N$
in the NLC for finding signals of the physics beyond  the SM.
The NLC with $\sqrt{s}=420$ GeV $e^-\gamma$ option will be able to probe
$\gamma WW$ interaction at the  level of the SM  quantum corrections.
We have also pointed out the possibility to  test
 the left-right symmetry of electroweak
interactions at the energies of the final phase of the NLC
through the same reaction. 
If the right-handed neutrino is light, this
reaction  offers a much better discovery reach for $W_2$ than the
pair production in $e^+e^-$ or $e^-e^-$ collisions.

\noindent{\bf Acknowledgements.}  We are indebted to Aarre Pietil\"a,
Turku, for useful discussions.  This work has been
supported by the Academy of Finland.

\newpage

\setlength{\unitlength}{5mm}

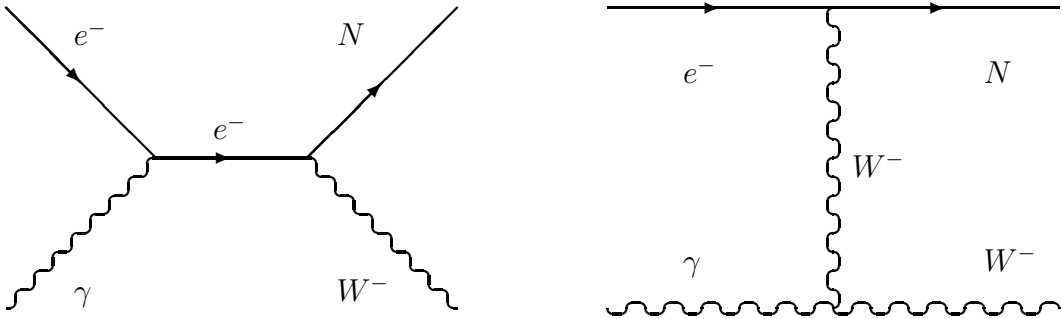
\begin{figure}

\begin{picture}(28,10)(0,-2)
\thicklines

\put(0,8){\vector(1,-1){2}}
\put(4,4){\line(-1,1){4}}
\put(4,4){\vector(1,0){2}}
\put(4,4){\line(1,0){4}}
\put(8,4){\vector(1,1){2}}
\put(8,4){\line(1,1){4}}

\multiput(3.5,3.75)(-0.5,-0.5){8}{\oval(0.5,0.5)[br]}
\multiput(4,3.75)(-0.5,-0.5){8}{\oval(0.5,0.5)[tl]}

\multiput(8.5,3.75)(0.5,-0.5){8}{\oval(0.5,0.5)[bl]}
\multiput(8,3.75)(0.5,-0.5){8}{\oval(0.5,0.5)[tr]}

\put(1.8,7){\shortstack{\(e^-\)}}
\put(1.8,0.2){\shortstack{\(\gamma\)}}
\put(8.8,7){\shortstack{\(N\)}}
\put(8.8,0.2){\shortstack{\(W^-\)}}

\put(5.5,4.5){\shortstack{\(e^-\)}}

\put(16,8){\line(1,0){12}}

\put(16,8){\vector(1,0){3}}
\put(16,8){\vector(1,0){9}}

\multiput(16.25,0.)(1,0){12}{\oval(0.5,0.3)[b]}
\multiput(16.75,0.)(1,0){12}{\oval(0.5,0.3)[t]}

\multiput(22.,7.25)(0,-1){8}{\oval(0.3,0.5)[r]}
\multiput(22.,7.75)(0,-1){8}{\oval(0.3,0.5)[l]}

\put(18,1){\shortstack{\(\gamma\)}}
\put(18,6){\shortstack{\(e^-\)}}
\put(26,1){\shortstack{\(W^-\)}}
\put(26,6){\shortstack{\(N\)}}
\put(22.5,3.5){\shortstack{\(W^- \)}}

\end{picture}

\caption{
Feynman diagrams for the  process $\protect e^-\gamma\rightarrow W^-N$.
}

\end{figure}

\begin{figure*}[hbtp]
\begin{center}
 \mbox{\epsfxsize=12cm\epsfysize=12cm\epsffile{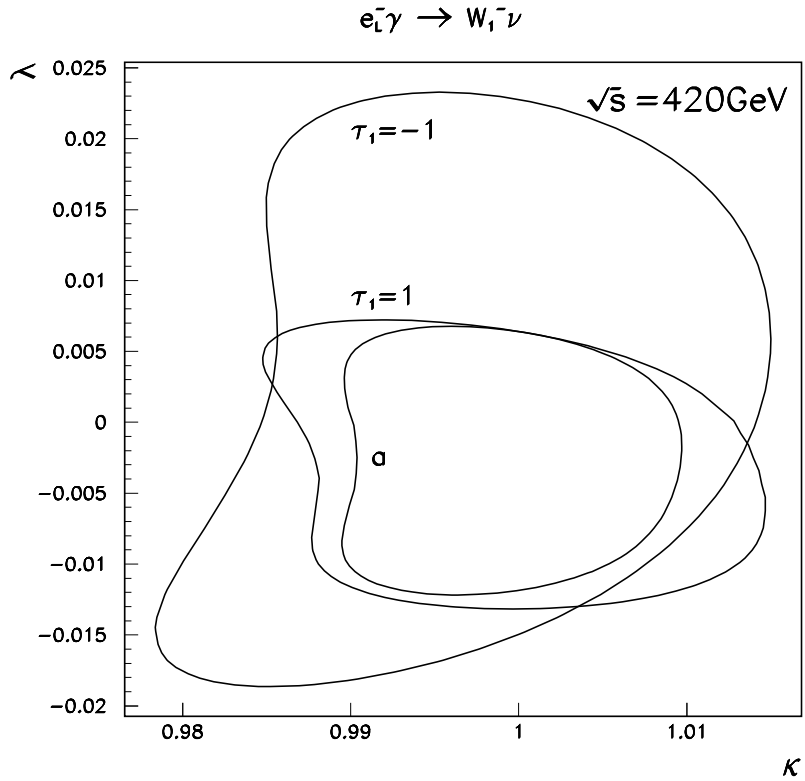}}
\caption{
The allowed domains of the photon anomalous coupling parameters
$\protect \kappa_{\gamma},\lambda_{\gamma}$ obtained by analysing 
the SM differential cross sections of  different photon polarization states
(as indicated on figure). The curve of combined analysis of the differential
cross sections is denoted by $a.$  
}
\end{center}
\end{figure*}

\begin{figure*}[hbtp]
\begin{center}
 \mbox{\epsfxsize=16cm\epsfysize=8cm\epsffile{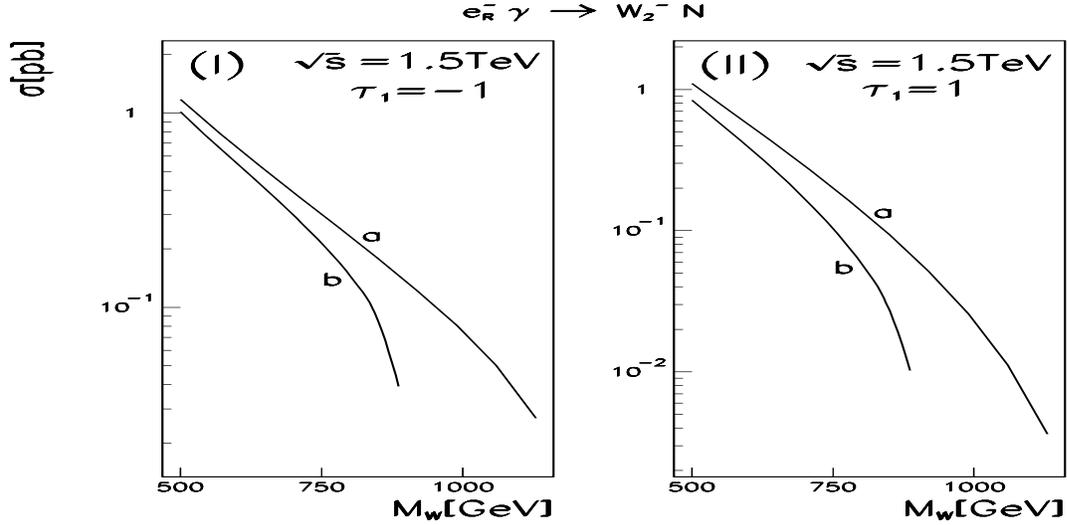}}
\caption{
The total cross section of the process
$\protect e_R^-\gamma\rightarrow W_{2}^-N $ 
as a function of heavy gauge boson mass for the left- (figure (I)) and
right-handedly (figure (II)) polarized photon beams.
The masses of heavy neutrino are taken to be 
 $\protect M_{N}=300$ GeV and $\protect M_{N}=600$
GeV for curves $a$ and $b,$ respectively. 
 }
\end{center}
\end{figure*}

\begin{figure*}[hbtp]
\begin{center}
 \mbox{\epsfxsize=12cm\epsfysize=8cm\epsffile{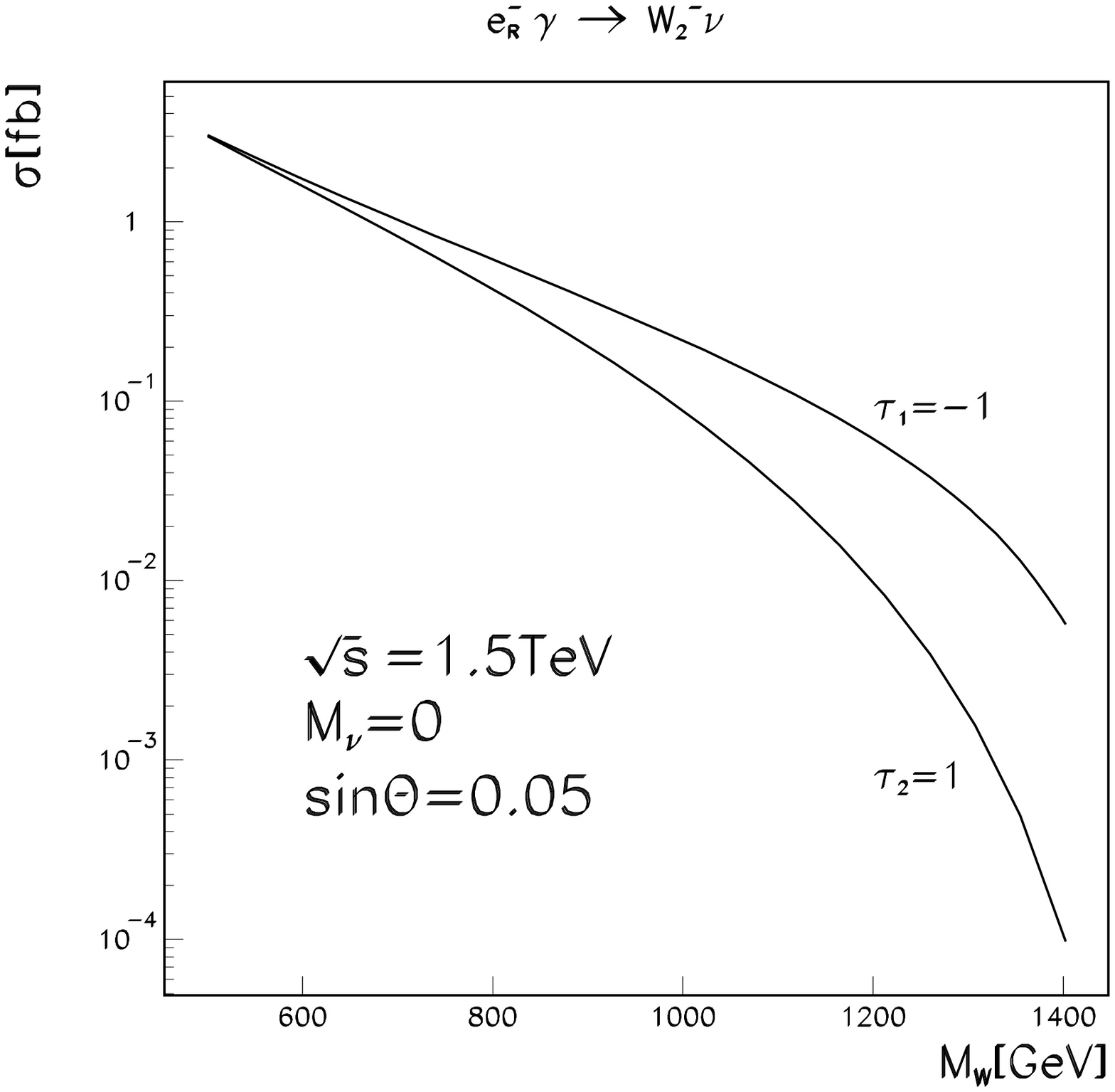}}
\caption{
The total cross section of the process
$\protect e_R^-\gamma\rightarrow W_{2}^-\nu $ 
as a function of heavy gauge boson mass for the left- and
right-handedly polarized photon beams.
The neutrino mixing angle is taken to be $\protect \sin\theta=0.05.$
 }
\end{center}
\end{figure*}


\begin{thebibliography}{99}

\bibitem{orava} 
See \eg, {\em Proc. of the Workshop on Physics and Experiments 
with Linear Colliders} 
(Saariselk\"a, Finland, September 1991), eds. R. Orava,
P. Eerola and M. Nordberg (World Scientific, 1992); \\ 
{\em Proc. of the Workshop on Physics and Experiments 
with Linear Colliders} 
(Waikoloa, Hawaii, April 1993), eds. F.A. Harris,
S.L. Olsen, S. Pakvasa and X. Tata (World Scientific, 1993).   

\bibitem{cuy} 
See \eg, 
P. Helde, K. Huitu, J. Maalampi and M. Raidal, 
Nucl. Phys. {\bf B 437} (1995) 305, and references therein.

\bibitem{ginz} 
I. Ginzburg, G. Kotkin, V. Serbo and V. Telnov, 
 Nucl. Instrum. Methods {\bf 205} (1983) 47.

\bibitem{telnov} V. Telnov, 
{\em Proc. of the First Arctic Workshop
on Future Physics and Accelerators} 
(Saariselk\"a, Finland,  1994),
eds. M. Chaichian, K. Huitu and R. Orava (World Scientific, 1995).

\bibitem{eg} 
A. Grau and J.A. Grifols, Nucl. Phys. {\bf B 233}(1984) 375.

\bibitem{choi} 
S.Y. Choi and F. Schrempp, Phys. Lett. {\bf B 272} (1991) 149.

\bibitem{king} 
K. Cheung, Nucl. Phys. {\bf B 403} (1993) 572.

\bibitem{lr} 
J.C. Pati and A. Salam, Phys. Rev. {\bf D 10} (1974) 275; \\
R.N. Mohapatra and J.C. Pati, Phys. Rev {\bf D 11} (1975) 566, 2558; \\
G. Senjanovic and R.N. Mohapatra, Phys. Rev. {\bf D 12} (1975) 1502.

\bibitem{mass} 
F. Abe {\em et al.}, Phys. Rev. Lett. {\bf 74} (1995) 2900.

\bibitem{hagi} 
K.J.F. Gaemers and G.J. Gounaris, Z. Phys. {\bf C 1} 
(1979) 159; \\
K. Hagiwara, R.D. Peccei, D.Zeppenfeld and K.Hikasa, 
Nucl. Phys. {\bf B 282}  (1987) 253. 

\bibitem{yehudai}
E. Yehudai, Phys. Rev. {\bf D 44} (1991) 3434.

\bibitem{phil} 
O. Phillipsen, Z. Phys. {\bf C 54} (1992) 643.

\bibitem{raidal} M. Raidal, 
Nucl. Phys. {\bf B 441 } (1995) 49.


\bibitem{sys} 
M. Frank, P.M\"atting, R. Settles and Z. Zeuner, in
{\em Proc. of the Workshop "$ e^+e^-$ Collisions at 500 GeV:
The Physics Potential,}" ed. P.M. Zerwas (1991), DESY 92-123B.

\bibitem{rad} J. Kodaira, H. Tochimura, Y. Yasui and I. Watanabe,
preprints HUPD-9509, OCHA-PP-57; \\
A. Arhrib, J.L. Kneur and G. Moultaka, preprint CERN-TH/95-344.

\bibitem{rizzoap95} T. Rizzo, Phys. Rev.  {\bf D 50} (1994) 325,
and $ibid.$ 5602.


%%%%%%%%%%%%%%%%%%%%




\end{thebibliography}
\end{document}